\documentclass[utf8]{FrontiersinHarvard} 

\usepackage{url,hyperref,lineno,microtype,subcaption}
\usepackage[onehalfspacing]{setspace}

\usepackage{ulem}
\usepackage{graphicx}
\usepackage{array}
\usepackage{multirow}

\usepackage{xcolor}
\usepackage{listings}
\usepackage{graphicx}
 
\textbf{}

\def\keyFont{\fontsize{8}{11}\helveticabold }
\def\firstAuthorLast{Louis, C. K. {et~al.}} 
\def\Authors{C. K. Louis\,$^{1,*}$, C. M. Jackman\,$^{1}$, S. W. Mangham\,$^{2}$, K. D. Smith\,$^{1}$, E. P. O'Dwyer\,$^{1}$, A. Empey\,$^{1}$, B. Cecconi\,$^{3}$, P. Zarka\,$^{3}$, S. Maloney\,$^{1}$}
\def\Address{$^{1}$School of Cosmic Physics, DIAS Dunsink Observatory, Dublin Institute for Advanced Studies, Dublin 15, Ireland\\
$^{2}$Department of Electronics \& Computer Science, University of Southampton\\
$^{3}$LESIA, Observatoire de Paris, CNRS, PSL Research University, Meudon, France}
\def\corrAuthor{Corresponding Author}
\def\corrEmail{corentin.louis@dias.ie}

\begin{document}
\onecolumn
\firstpage{1}

\title[The SPACE Labelling Tool]{The ``SPectrogram Analysis and Cataloguing Environment'' (SPACE) Labelling Tool} 

\author[\firstAuthorLast ]{\Authors} 
\address{} 
\correspondance{} 

\extraAuth{}

\maketitle

\begin{abstract}

The SPectrogram Analysis and Cataloguing Environment (SPACE) tool is an interactive python tool designed to label radio emission features of interest in a time-frequency map (called ``dynamic spectrum''). The program uses Matplotlib's Polygon Selector widget to allow a user to select and edit an undefined number of vertices on top of the dynamic spectrum before closing the shape (polygon). Multiple polygons may be drawn on any spectrum, and the feature name along with the coordinates for each polygon vertex are saved into a ``.json'' file as per the ``Time-Frequency Catalogue'' (TFCat) format along with other data such as the feature id, observer name, and data units. This paper describes the first official stable release (version 2.0) of the tool.


\tiny
 \keyFont{ \section{Keywords:} Python tool, Labelling tool, Planetary radio emissions, Solar radio emissions, Jupiter, Saturn, Earth, Sun} 
\end{abstract}

\section{Introduction}

Non-thermal planetary radio emissions are produced by out-of-equilibrium populations of charged particles in planetary magnetospheres, and are observed at almost all strongly magnetized planets in our solar system: the Earth, Jupiter, Saturn, Uranus and Neptune. The radio emissions can be divided into different classes, such as plasma waves, electromagnetic radio waves or electrostatic radio waves. It is highly desirable to select these distinct classes, which can often have characteristic frequency ranges, morphologies, or polarizations. Once catalogues of different emission types have been built up, that can enable large statistical studies unveiling both the average and extreme behaviour of planetary radio emissions.

There have been many long-running planetary spacecraft which have returned huge volumes of radio data and we have only scratched the surface of its analysis. For example, the Cassini mission at Saturn spent 13 years studying the kronian system, revealing several components to its radio spectrum \citep{2008JGRA..113.7201L, 2011pre7.conf...99Y, 2017pre8.conf..171L, 2011pre7.conf..115T}. Furthermore, the Wind spacecraft has spent almost two decades observing terrestrial (and solar) radio emissions from a range of vantage points near Earth \citep{2021JGRA..12629425W, 2022JGRA..12730209F, 2008A&A...489..419B}.

Significant efforts have been made in recent years to classify radio emissions from Jupiter, where the non-thermal radio emission is composed of half a dozen components \citep{2021JGRA..12629435L}. These components overlap themselves in time and frequency, making automatic detection non-trivial. Therefore, manually cataloguing them is mandatory to be able to study them independently. Previous catalogues have been using square boxes to define the time and frequency intervals containing the radio signal \cite[such as][]{2020_Leblanc_Catalogue_1, 2020_Leblanc_Catalogue_2, 2020_Leblanc_Catalogue_3, 2020_Leblanc_Catalogue_4, 2020_Leblanc_Catalogue_5}, but to be able to automatically disentangle the emissions when using the catalogue, it is then needed to construct catalogues with polygon vertices and with distinct labels. This was done by a few authors, primarily using tools built in IDL to construct such catalogues of Jupiter radio emissions \cite{2017A&A...604A..17M, 2021JGRA..12629780Z}. Once catalogues are built they can comprise training sets which form the basis of supervised machine learning approaches to classify larger samples of unseen data.


Here we present a Python user interface tool to allow the drawing of polygons around features in dynamic spectra. Section \ref{sec:package} describes the package, and an example of the use of this package. 
Section \ref{sec:history} summarises the version history of the code.
Section \ref{sec:application} gives examples of the application of the catalogues produced by the tool. Section \ref{sec:improvements} presents some future avenues to continue to improve the tool.

\section{The SPACE package} \label{sec:package}

The SPectrogram Analysis and Cataloguing Environment (SPACE) tool is an interactive python tool designed to label radio emission features of interest in a time-frequency map (called a ``dynamic spectrum''). The program enables users to create and edit the vertices of a polygon on the dynamic spectrum plot, before naming and saving it as a 'feature' in a catalogue for future analysis.

\subsection{Installation}

The code is available on an open-source GitHub repository \citep{2022_SPACE_code_Louis}, with full installation instructions present on the repository page, and packed with the tool. It can easily be downloaded and installed using \texttt{git} and \texttt{pip}, with all prerequisites included in the provided \texttt{requirements.txt} file.

\subsection{Usage}

Once installed using \texttt{pip}, the space labelling tool is available as a system-wide command \texttt{spacelabel}. It can then be used to view and label spacecraft observational data, by providing an input file in \textbf{HDF5} or \textbf{NASA CDF} format (see section \ref{sec:usage:input} for specifics), and a time window to view. An example use is shown in Figure \ref{fig:figure_1}.

Users may select any number of the measurement types present in the file (e.g. polarisation, flux and/or power), and view them all tiled on the screen. For example, Figure \ref{fig:figure_1}a displays Cassini Flux and Polarization radio data, while Figure \ref{fig:figure_1}b only display Juno Flux radio data. Users can then click to select polygonal regions of the observation to label as features (top panel of Figure \ref{fig:figure_1}a). To close the polygon, users should click on the first drawn vertex of the polygon. Once done, a window pops up and ask to name the drawn feature appropriately (see top panel of Figure \ref{fig:figure_1}a). Features labelled in one view (e.g. intensity) appear simultaneously on the other views once they have been named (see bottom panel of Figure \ref{fig:figure_1}a), allowing users to easily see how a feature presents in multiple measurement types.

Once a region has been labelled, a user can pan their viewing window back and forth through the time range within the dataset by clicking on the \texttt{Prev} or \texttt{Next} buttons (see Figure \ref{fig:figure_1}), with an overlap applied between each view in order to facilitate labelling features that lie on the edge of a window. Once finished, the labelled regions can be saved as a \textbf{TFCat} (Time-Frequency Catalogue) formatted  \textit{JSON}\footnote{\url{https://www.json.org/}} file (Cecconi et al., 2022, this issue) by clicking on the \texttt{Save} button, and used later. If a user re-opens the same data file, or another data file with the same naming structure (e.g. \texttt{observations\_20180601\_v02.cdf} and \texttt{observations\_20180602\_v02.cdf}) saved features from previous sessions will be pre-loaded (see Figure \ref{fig:figure_1}).

\begin{figure*}
    \centering
    \includegraphics[width=0.85\textwidth]{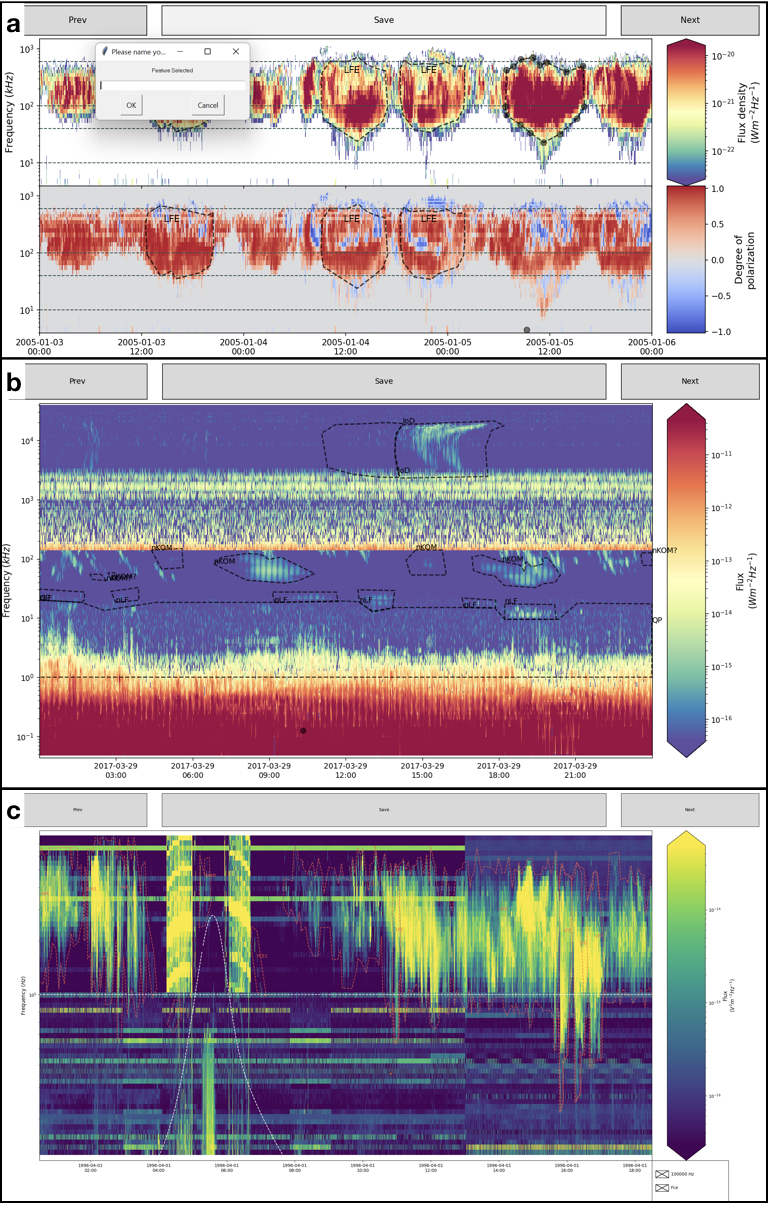}
    \begin{center}
        {\textbf{Figure 1}: Caption on next page}
    \end{center}
\label{fig:figure_1}
\end{figure*}
\begin{figure*}
    \centering
    \caption{Examples of plots from the SPACE labelling Tool.\newline
    Panel (a) displays Cassini/RPWS \citep[Radio and Plasma Waves Science][]{2004SSRv..114..395G} data \citep{2008JGRA..113.7201L, 2009Cassini_RPWS_LESIA_data_collection_Lamy}. The two panels show Intensity and Polarization data, respectively. At the top right of the top panel one can see a polygon that has just been drawn, with the window for naming the feature appearing at the top left of the graphics window. Other features have already been labelled, and appear in both intensity and polarisation views, with their names overlaid.\newline
    The data displays in panel (b) are the estimated flux density \citep{2021Juno_Waves_Calibrated_data_collection, 2021JGRA..12629435L} from by Juno/Waves measurements \citep{2017SSRv..213..347K}, with the \cite{2021_Juno_Waves_catalog} catalogue overlaid.\newline
    Panel (c) displays observations of Polar/PWI instrument \citep{1995SSRv...71..597G}. The horizontal dashed-white line shows an example of the use of the \texttt{-horizontal\_line} option. The variable dashed-white line show that the tool is also able to read a 1D table from the CDF file (provided that this has been specified in the configuration file)}
\label{fig:figure_1}
\end{figure*}

Full usage documentation is available on the GitHub repository for the code \citep{2022_SPACE_code_Louis}.

\subsubsection{Procedure}

When the code first opens a datafile, it compares the columns within to a selection of pre-made (and user-creatable) 'configuration' files for each type of input file (e.g. \textbf{CDF}, \textbf{HDF5}). Each describes a file in terms of the column names within it, and provides metadata for use in the tool - units and display names, and scaling factors that can be applied to change data stored in one unit into data viewed in another. If the data are in an unspecified format, the code will prompt the user to create a configuration file (see section \ref{sec:usage:input}). Alternatively, if their data fit multiple configuration files, they will be prompted to select which they want to use (see section  \ref{sec:usage:options}).

Once a configuration has been determined, the code parses the observations and may re-bin the time into a coarser resolution in order to improve performance, taking the average of measurements in the new larger bins. It can also rescale the frequency data into evenly-spaced logarithmic bins between the minimum and maximum bins in the original data, as some data files have non-monotonic bin structures. When altering the frequency bins, measurements are logarithmically interpolated between the readings on the previous scale. Default adjustments can be defined in configuration files for file types, and may be over-ridden by command-line arguments to the tool. The parsed and adjusted data are then re-saved as a compressed \textbf{HDF5} file, reducing both the size of the data and time to access.

The pre-processed data are then displayed to the screen using MatPlotLib \citep{Hunter:2007}, in a window the size of the user's initial time range. The dynamic range of the data is constrained to improve visibility of features; displaying, by default, the 5th-95th quantiles of the signal for each measurement (to prevent anomalously low or high values overly-compressing the ranges of interest, reducing the ability to discern features). GUI buttons allow users to pan between time windows of equivalent size - with each window will overlap the previous window by 25\% in the direction of travel. Features can then be defined by drawing polygons on the time window using the MatPlotLib Polygon Selector widget.

\subsubsection{Options} \label{sec:usage:options}

The user must specify the path to the file they want to visualise (or 'first' file in a collection, e.g. of \textbf{CDF} files), along with the start and end dates of their initial viewing window, in \textbf{ISO} \textit{year-month-day hour:minute:second} format (e.g. \texttt{2018-06-12 18:00:01}). However, there are further options available:

\begin{itemize}
    \item \texttt{-f FREQUENCY\_BINS}: Rescales the data to this many evenly-spaced logarithmic bins. Overrides any default set in the configuration files.
    \item \texttt{-t MINIMUM\_TIME\_BIN}: Rebins the data to time bins of this size, if it is currently more finely binned. The bin size need to be given in second 
    \item \texttt{-s SPACECRAFT}: Specifies the name of the spacecraft configuration file to use, if multiple describe the datafile the user has provided. 
    \item \texttt{-frac\_dyn\_range FRAC\_MIN FRAC\_MAX}: Defines the dynamic range of the colour bar in the visualisation, as a fraction of the distribution of values in the data file. This must be numbers between $0$ and $1$. Default values are $0.05$ and $0.95$ (the 5th-95th quantiles of the displayed signal)
    \item \texttt{-cmap CMAP}: The name of the color map that will be used for the intensity plot (by default: viridis)
    \item \texttt{-g VALUES\_1 ... VALUES\_N}: Draws horizontal line(s) on the visualisation at these specified frequencies to aid in interpretation of the plot. Values must be in the same units as the data. Lines can be toggled using check boxes.
    \item \texttt{--not\_verbose}: If \texttt{not\_verbose} is called, the debug log will not be printed. By default: verbose mode
\end{itemize}
    
\subsubsection{Input Formats} \label{sec:usage:input}

The code is designed to cope with input files in a variety of file formats and column formats by use of configuration files, several of which are pre-provided. \textbf{HDF5} input files require at least three datasets, corresponding to \textbf{observation time} (floats, in MJD), \textbf{frequency range} (floats, in any arbitrary unit) and at least one measurement, stored with frequency as the rows and observation as the columns. The names and units for each measurement (in LaTeX form) must be provided in a configuration file, in easily-editable \textbf{JSON} format. The appropriate configuration files are automatically-selected by the code from those available - making it easy to work with \textbf{HDF5} files from a variety of collaborators with arbitrary naming schemes.

\textbf{CDF} files in NASA format are more structured, and can be read in either singly or as a collection, combining all files in the directory matching the naming scheme \texttt{[...]\_YYYMMDD\_[...].cdf} into a single pre-processed data file. As with the \textbf{HDF5} files, \textbf{CDF} files must contain a frequency attribute (floats, in any arbitrary unit) and a time attribute \citep[either in \texttt{TT\_2000} or \texttt{CDF\_EPOCH} format, which is parsed using \textbf{Astropy},][]{price2018astropy} and at least one measurement, stored with frequency as the rows and observation as the columns.

The code can easily be expanded to ingest other file formats (see section \ref{sec:structure:input}).

\subsection{Structure}

\subsubsection{'Model-View-Presenter' Architecture}
The code is designed using a standard object-oriented 'Model-View-Presenter' architecture, with strong separation between the data input and management, and the visualisation. This allows for easy development of both new input file-types (see section \ref{sec:structure:input}) and pre-processing options, and alternative GUI front-ends and settings (see section \ref{sec:improvements} for suggested development building on top of this flexibility). A generic 'Presenter' controls the logic of the program, and feeds data from the data models to the selected GUI view, and requests from the GUI for changes to the data models. Either the 'View' or 'Model' can be easily interchanged as long as they conform to the API expected by the 'Presenter'.

Full development documentation is available on the GitHub repository for the code \citep{2022_SPACE_code_Louis}.

\subsubsection{Addition of new Input Formats} \label{sec:structure:input}

New input formats can be easily added by extending the base \texttt{DataSet} class included in the code. A developer only needs to define the routines for inputting the data from file; the code will then handle pre-processing and data access.

\section{History of the code}
\label{sec:history}

The first version of the labelling code was developed in IDL by P. Zarka. It allowed users to read data from an IDL saveset (\texttt{sav} format), draw polygons around features of interest and label them. However, this IDL version had to be adapted to each new dataset. This code has been used to build many catalogues based on different observers (such as the Nançay Decameter Array (NDA) ground-based radio telescope \citep{2017A&A...604A..17M}, or the Cassini \citep{2021JGRA..12629780Z} or Juno \citep{2021JGRA..12629435L, 2021_Juno_Waves_catalog} spacecraft).

The second version of the labelling tool was written in Python and was the first to be officially released \citep{2021_SPACE_code_Empey}. This version allowed to automatically read any dataset in \texttt{sav} or \texttt{cdf} format, based on the information requested from users from the terminal. The other main improvements compared to the previous version were the number of vertices in the polygons (unlimited) and the possibility to modify the vertices position during the polygon drawing (using the \texttt{Matplotlib's Polygon Selector widget}), as well as the production of the catalogue directly in TFCat format.

The current version \citep{2022_SPACE_code_Louis} brings a large number of improvements, both in terms of architecture, usability and ergonomics, which are described in the previous sections.

\section{Applications} \label{sec:application}

Once a catalogue has been produced, it can also be displayed using the SPACE labelling Tool (see Figure \ref{fig:figure_1} or the Autoplot Software \citep{2010sdfghF}. For an example, the reader is invited to visit the web page \url{https://doi.org/10.25935/nhb2-wy29} where an autoplot template file is given to display the Juno data \citep{2021Juno_Waves_Calibrated_data_collection} and the \cite{2021_Juno_Waves_catalog} catalogue overlaid in autoplot. See Cecconi et al. (2022, this issue) for more information about the display of a Catalogue using Autoplot.
The catalogue can be used to study the different components of the radio emission spectrum, e.g. as done by \cite{2021JGRA..12629435L}, where the data can then be automatically selected using the catalogue via a mask or an inverse-mask. In the case presented in Figure \ref{fig:figure_1}, not every type of emission is labelled, but in each frequency range (kilometric or hectometric) only one radio component remains.  We can then study the components one by one (e.g. their latitudinal distribution, as in \cite{2021JGRA..12629435L}, their distribution as a function of observer’s or Sun’s longitude, as in \cite{2021JGRA..12629780Z},or their distribution versus observer’s longitude and satellite (Io) phase as in \cite{2017A&A...604A..17M}).

With the SPACE labelling Tool, we are also providing some useful routines to use the catalogue\footnote{\url{https://github.com/elodwyer1/Functions-for-SPACE-Labelling-Tool}}.

These catalogues can also then be used to train machine learning algorithm to detect automatically the radio emissions in past (Cassini, NDA) or future observation (such as Juno, JUICE, NDA).

\section{Limitations \& Future Work} \label{sec:improvements}

The code is ready for distribution and use, but has some technical limitations. Potential works to address those limitations, and avenues of future development, are:

\begin{itemize}
        \item \textbf{Performance:} The MatPlotLib-based front-end can struggle when provided with especially high resolutions of data, or over large time windows. Rebinning features exist to mitigate this, but ideally the front-end would be re-implemented in a more performant framework (e.g. Plotly \citet{plotly}).
        \item \textbf{Scalability:} The code loads all the data provided into memory at launch, limiting its applicability for large datasets. Whilst this can be mitigated by the feature to allow appending to \textbf{TFCat} files created by data files sharing filename formats, a 'deferred load' approach would be better. This would be best accomplished using the Dask and XArray libraries \citep{dask, hoyer2017xarray}.
        \item \textbf{Configurations:} The code depends heavily on pre-written configuration files, and can prompt users to create missing ones - but does not yet contain a 'wizard' or automatic walkthrough to aid users in creating them.
        \item \textbf{Catalogue integration:} The modular format of the code would make it possible to create 'dataset' types that access and download data directly from online catalogues, maintaining local caches.
        
\end{itemize}

\section*{Conflict of Interest Statement}

The authors declare that the research was conducted in the absence of any commercial or financial relationships that could be construed as a potential conflict of interest.

\section*{Author Contributions}
C. K. Louis and C. M. Jackman led the development of the SPACE Labelling Tool. C. K. Louis wrote the paper and worked on the development of the code. P. Zarka developed the first IDL version of the code. A. Empey developed the first python version of the code. S. Maloney worked on the development of the second version of the code. S. W. Mangham developed the current version of the code.
E. P. O'Dwyer tested the different versions of the code and give inputs to the developers of the code. K. D. Smith added features to the latest version of the code. B. Cecconi led the TFCat format of the catalogues.  All the co-authors have contributed to the writing of the paper.

\section*{Funding}
C.K. Louis', C.M. Jackman's, E. P. O'Dwyer's and A. Empey's work at the Dublin Institute for Advanced Studies was funded by Science Foundation Ireland Grant 18/FRL/6199. S. Mangham's work at the University of Southampton Research Software Group was funded by a grant via the Alan Turing Institute. K.D. Smith's work at the Dublin Institute for Advanced Studies was funded by a 2022 SCOSTEP/PRESTO Grant.



\section*{Data Availability Statement}
The code of the SPACE labelling Tool is open-source and freely available on github \citep{2022_SPACE_code_Louis}. The Cassini/RPWS dataset displayed in Figure \ref{fig:figure_1}a, produced by \citet{ 2008JGRA..113.7201L} is available at \url{https://doi.org/10.25935/ZKXB-6C84} \citep{2009Cassini_RPWS_LESIA_data_collection_Lamy}. The Juno/Waves dataset displayed in Figure \ref{fig:figure_1}b, produced by \citet{2021JGRA..12629435L}, is accessible at \url{https://doi.org/10.25935/6jg4-mk86} \citep{2021Juno_Waves_Calibrated_data_collection}, and the catalogue can be download at \url{https://doi.org/10.25935/nhb2-wy29} \citep{2021_Juno_Waves_catalog} . The Polar/PWI dataset displayed in Figure \ref{fig:figure_1}c is accessible through the CDAWeb at \url{https://cdaweb.gsfc.nasa.gov/pub/data/polar/pwi/}.

\bibliographystyle{Frontiers-Harvard} 
\bibliography{bibliography}


\section*{Figure captions}


\end{document}